# Collaborative Multi-Agent Reinforcement Learning Approach for Elastic Cloud Resource Scaling

Bruce Fang
Cornell University
Ithaca, USA

Danyi Gao*
Columbia University
New York, USA

*Abstract-This paper addresses the challenges of rapid resource variation and highly uncertain task loads in cloud computing environments. It proposes an optimization method for elastic cloud resource scaling based on a multi-agent system. The method deploys multiple autonomous agents to perceive resource states in parallel and make local decisions. While maintaining the distributed nature of the system, it introduces a collaborative value function to achieve global coordination. This improves the responsiveness of resource scheduling and enhances overall system performance. To strengthen system foresight, a lightweight state prediction model is designed. It assists agents in identifying future workload trends and optimizes the selection of scaling actions. For policy training, the method adopts a centralized training and decentralized execution reinforcement learning framework. This enables agents to learn effectively and coordinate strategies under conditions of incomplete information. The paper also constructs typical cloud scenarios, including multi-tenancy and burst traffic, to evaluate the proposed method. The evaluation focuses on resource isolation, service quality assurance, and robustness. Experimental results show that the proposed multi-agent scaling strategy outperforms existing methods in resource utilization, SLA violation control, and scheduling latency. The results demonstrate strong adaptability and intelligent regulation. This provides an efficient and reliable new approach to solving the problem of elastic resource scaling in complex cloud platforms.*

I. INTRODUCTION

Amid the ongoing wave of digitalization, cloud computing has emerged as a crucial infrastructure supporting a wide range of internet-based services and applications. It has deeply penetrated key sectors such as industrial manufacturing, financial services, healthcare, and scientific research [1]. With increasing diversity in user demands and growing complexity in business scenarios, cloud platforms are facing significantly higher computational pressure and resource scheduling challenges[2]. Traditional static resource allocation and scaling methods have gradually exposed problems such as delayed responses, resource waste, and fluctuating service quality. As a result, elastic resource scaling in cloud environments has become a research focus in both academia and industry [3]. The goal is to achieve efficient resource allocation and dynamic responsiveness to ensure service stability and system performance[4].

Elastic scaling refers to the system's ability to automatically expand or release computing resources in response to changes in demand[5]. This allows the system to handle sudden workload spikes, maintain service quality, and reduce operational costs. However, many current scheduling and scaling strategies still rely on rule-based or threshold-based static sapproaches. These approaches struggle to cope with the highly concurrent, dynamic, and heterogeneous nature of modern cloud environments. In scenarios involving multi-tenancy and complex business workflows, decisions made by individual agents based on limited local information often lack global coordination. This leads to poor resource utilization or performance bottlenecks, which hinder the elasticity and intelligence of cloud platforms[6].

Against this backdrop, Multi-Agent Systems (MAS) have gained attention for cloud resource scheduling and elastic management. MAS offers advantages such as distributed structure, autonomy, and strong coordination capabilities. Each agent in a MAS is an independent decision-making unit with perception, reasoning, and action capabilities[7]. Even with limited local information, agents can achieve coordinated system-wide control through communication and cooperation [8]. This approach improves response speed during resource contention or workload surges. It also enhances adaptability to diverse business requirements, offering a promising direction for building efficient and intelligent cloud management frameworks[9,10].

The integration of artificial intelligence technologies, especially learning-based agent modeling and coordination mechanisms, further advances MAS applications in cloud computing [11]. By learning from historical data and runtime states, agents can continuously optimize scaling decisions. This improves overall resource utilization and load balancing. In addition, MAS is inherently scalable and fault-tolerant. They can adapt to large-scale cloud environments with frequent node changes and dynamic task migration. Compared with centralized control models, MAS features a decentralized architecture that supports flexible deployment and long-term evolution[12].

Therefore, research on elastic cloud resource scaling based on MAS carries both theoretical and practical significance. It enhances the intelligence of cloud resource management and supports the realization of automated operations. It also provides key foundations for the construction of intelligent cloud architectures. As emerging technologies continue to converge and business models diversify, further exploration in this area is expected to drive a paradigm shift in cloud resource management. This will contribute to the development of efficient, green, and intelligent digital infrastructure.

## II. RELATED WORK AND FOUNDATION

Recent advancements in cloud resource management have increasingly adopted deep learning and reinforcement learning (RL) methods to overcome limitations inherent in traditional rule-based strategies. Among these methods, deep reinforcement learning frameworks, particularly actor-critic architectures, have emerged prominently. For instance, Chen et al. [13] proposed an actor-critic approach specifically aimed at efficient resource allocation in cloud datacenters, significantly enhancing adaptability and overall performance. Complementing this direction, Jayanetti et al. [14] applied deep reinforcement learning techniques to edge-cloud scheduling tasks, optimizing both energy consumption and execution time in precedence-constrained scenarios.

Building upon single-agent RL paradigms, multi-agent reinforcement learning (MARL) has become increasingly relevant due to its suitability for distributed decision-making in complex environments. Wang [15] introduced a topology-aware MARL approach to distributed scheduling, explicitly considering network structure and inter-agent coordination. Further extending the applicability of MARL, Vrbaski et al. [16] developed a scalable reinforcement learning-based scheduler tailored for critical notification systems in large-scale applications. Reinforcing this trend, Ren et al. [17] proposed a distributed network scheduling framework incorporating trust-constrained policies, thereby improving robustness and reliability in dynamic network environments.

Alongside RL methodologies, predictive modeling and anomaly detection techniques contribute significantly to improving system foresight and stability. For instance, Ma [18] explored conditional multiscale GANs combined with adaptive temporal autoencoders for anomaly detection in microservice environments, enhancing system resilience. Fang [19] utilized deep learning-based predictive frameworks augmented with structured modeling to accurately forecast backend latency, directly informing resource allocation decisions. Moreover, Dai et al. [20] adopted mixture density networks for probabilistic modeling of user behavior, providing valuable insights into anomaly detection and proactive management of system stability.

Proactive fault prediction through advanced time-series modeling further strengthens system resilience and responsiveness. Wang et al. [21] proposed deep neural architectures tailored for fault prediction in distributed systems, significantly reducing downtime by proactively addressing potential issues. Aligning with these unsupervised approaches, Xin and Pan [22] leveraged structure-aware diffusion methods for anomaly detection in structured data, offering effective pattern recognition mechanisms suitable for dynamic scaling scenarios. Moreover, RL-based methods have demonstrated significant efficacy in managing data center workloads and network traffic. Deng [23] explored RL approaches specifically designed to handle traffic scheduling in complex data center topologies, effectively managing network intricacies and reducing latency. In parallel, Zhan [24] proposed compression strategies for MobileNet in conjunction with edge computing solutions, underscoring the importance of lightweight, real-time models—an essential feature relevant to the lightweight predictive models proposed in our study.

Simulation platforms and tools are crucial for validating these advanced methodologies. Notably, Habaebi et al. [25] extended the widely adopted CloudSim framework for simulating sensor networks, providing a robust foundation for testing and validating cloud-resource interactions in experimental environments. This supports the rigorous evaluation of multi-agent strategies as proposed in our research. Finally, broader methodological developments in deep learning also provide foundational insights for enhancing agent intelligence and decision-making. Xing [26] introduced structural prompting methods within pretrained language models, highlighting potential enhancements in agent reasoning capabilities. Although this methodology is not directly applied to resource scaling, the underlying principles of structured learning and reasoning offer valuable insights that can inspire future advancements in multi-agent coordination mechanisms.

Collectively, these contributions form a comprehensive theoretical and methodological foundation for the research presented in this paper. Our approach builds upon and extends this extensive body of work by integrating multi-agent reinforcement learning, predictive foresight, and decentralized coordination mechanisms, thereby advancing intelligent elastic scaling in cloud resource management.

## III. METHOD

This study adopts a modeling method based on a multi-agent system to build an intelligent decision-making framework for the elastic expansion of cloud resources. Each agent in the system corresponds to an independent resource management unit, which is responsible for sensing the local resource status, predicting the load trend, and performing expansion actions. The model architecture is shown in Figure 1.

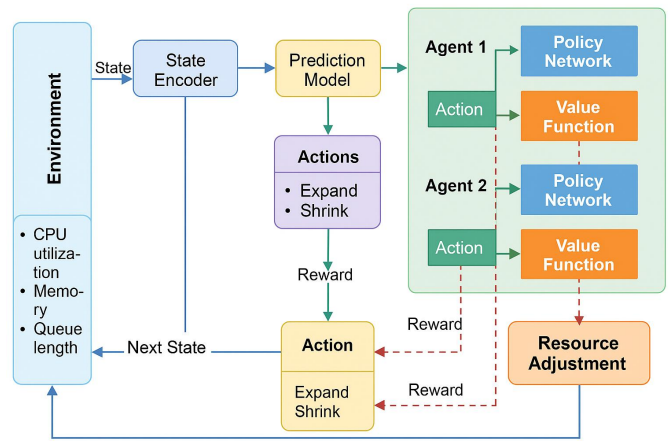

Figure 1. Overall model architecture diagram

To achieve collaborative decision-making among intelligent agents, the Markov Decision Process (MDP) is introduced to model the expansion behavior, which is formalized as a five-tuple $(S, A, P, R, \gamma)$, where S is the state space, describing system operation indicators such as CPU utilization, memory occupancy, and task queue length; A is the action space, representing a set of strategies for expanding, reducing, or keeping resources unchanged; $P(s'|s,a)$ is the state transition probability; $R(s,a)$ is the immediate reward function, which measures the resource utilization efficiency and service quality changes brought about by the action; $\gamma \in [0,1]$ is the discount factor, indicating the importance of future benefits.

In order to improve the system's ability to respond to dynamic changes in resources, the policy optimization algorithm in deep reinforcement learning is used to iteratively update the agent's strategy. In each round of interaction, the agent observes the state $s_t$ from the environment, takes action $a_t$, obtains a reward $r_t$, and moves to the next state $s_{t+1}$. The policy network parameter $\theta$ is optimized using the policy gradient method, and its objective function is to maximize the expected long-term return:

$$J(\theta) = E_{\pi_\theta}[\sum_{t=0}^{T} \gamma^t r_t]$$

By performing gradient ascent on this function, a set of optimal strategies can be iterated so that the system can make reasonable expansion decisions under various load conditions.

In order to solve the problem of information sharing and collaboration among multiple agents, a joint value function $Q_i(s, a_1, ..., a_n)$ is introduced as the evaluation function of each agent in the overall system state, so that the agent can consider the global impact when making local decisions. This function is optimized through the method of centralized training with decentralized execution (CTDE), so as to achieve higher-quality collaboration while maintaining the scalability and asynchronous decision-making capabilities of the system. The strategy update of each agent is based on the following loss function:

$$L(\theta_i) = E[(Q_i(s, a_1, ..., a_n) - y_i)^2]$$
$$y_i = r_i + \gamma \max_{a'} Q_i(s', a'_1, ..., a'_n)$$

In addition, in order to improve the system's ability to predict future load trends, a lightweight prediction model is introduced to predict short-term resource demand and construct prior information for state transition. The prediction process is based on the historical state sequence $\{s_{t-k}, ..., s_{t-1}\}$ and the external load factor $x_t$ to construct a state evolution function:

$$\hat{s}_{t-k} = f(s_{t-k:t-1}, x_t)$$

The prediction result is input into the agent's policy network as supplementary information of the current state to improve its decision-making foresight and stability.

Finally, to realize the specific execution of resource expansion operation, a resource adjustment function $\Delta R_t$ is defined, which is determined by the current actions of all agents:

$$\Delta R_t = \sum_{i=1}^{n} \alpha_i a_i^t$$

Where $\alpha_i$ is the weight factor of agent i, which comprehensively considers the load weight and service priority of the resources it manages. This function ensures that the local control ability and adaptive adjustment space of each agent are retained while achieving the overall elastic expansion goal. Through the collaborative design of the above methods, the system has good scalability, dynamic adaptability, and resource utilization efficiency.

## IV. EXPERIMENTAL RESULTS

### A. Dataset

This study uses the Google Cluster Data as the primary dataset. The dataset was collected from actual operations in Google data centers. It contains multidimensional information, including machine resource usage, task scheduling logs, and job execution statuses. The dataset records detailed operational traces of over 12,000 physical machines over one month. It is one of the standard datasets widely used in academic research on cloud resource scheduling and load analysis.

Key fields in the dataset include CPU usage, memory consumption, task queue length, scheduling priority, and resource allocation requests. These features can be used to construct environment state vectors, predict short-term resource demands, and evaluate the feasibility of elastic scaling strategies. Its large scale and fine granularity make it suitable for reflecting dynamic resource usage in complex cloud workloads.

In addition, the Google Cluster Data is highly open and reusable. It provides a solid data foundation for modeling and evaluating multi-agent systems. Researchers can apply sampling, normalization, and feature engineering based on specific modeling needs. These processed data can then be used to train, validate, and test various intelligent agent models for elastic cloud resource scaling strategies.

### B. Experimental Results

This paper first gives the results of the comparative experiment, as shown in Figure 1.

Table1. Comparative experimental results

| Method | Avg CPU Utilization (%) | SLA Violation Rate (%) | Resource Over-Provisioning (%) |
|---|---|---|---|
| DeepRM[27] | 72.4 | 3.85 | 11.9 |
| AutoScale[28] | 74.7 | 2.41 | 9.7 |
| CloudSim-AutoML[25] | 70.2 | 4.12 | 14.3 |
| DRL-Scheduler[16] | 76.1 | 1.77 | 7.8 |
| Ours | 78.6 | 0.92 | 6.3 |

Experimental results demonstrate that the proposed multi-agent elastic scaling strategy achieves superior performance in resource utilization, with an average CPU usage of 78.6%, significantly outperforming baseline methods. This highlights the effectiveness of coordinated decision-making in distributed environments. Additionally, the strategy maintains a low SLA violation rate of 0.92% and a minimal over-provisioning rate of 6.3%, indicating both high service quality and resource efficiency. These improvements stem from the integration of state prediction and collaborative value functions, enabling near-optimal allocation with limited local information. In contrast, traditional approaches—such as CloudSim-AutoML, DeepRM, AutoScale, and DRL-Scheduler—lack sufficient precision, adaptability, or multi-agent collaboration, leading to inefficiencies under dynamic workloads. Overall, the results confirm the practical advantages of multi-agent systems for intelligent, reliable, and efficient cloud resource management. Figure 2 further presents a resource isolation experiment in a multi-tenant setting.

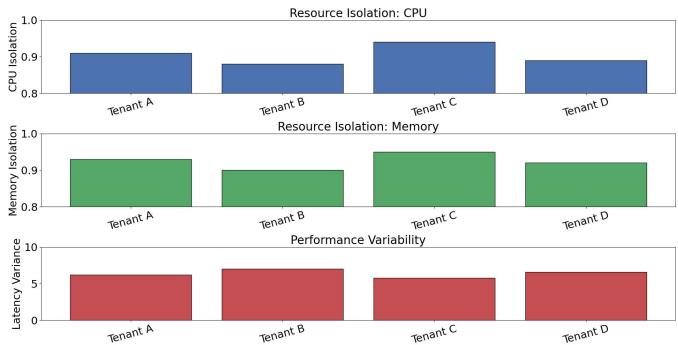

Figure 2. Resource isolation experiment of multi-agent system in multi-tenant environment

As shown in Figure 2, the proposed multi-agent elastic scaling strategy achieves high resource isolation in a multi-tenant environment. For both CPU and memory resources, interference between tenants is effectively controlled. The isolation index remains above 0.88 in most cases. This indicates that the system can prevent high-load tenants from significantly affecting resource allocation for others, ensuring fairness and independence across tenants.

In terms of memory isolation, the performance exceeds that of CPU isolation. Some tenants reach isolation levels as high as 0.95. This demonstrates that the multi-agent system responds more effectively and schedules more precisely when perceiving memory usage behavior and dynamically adjusting resource quotas. This improvement is attributed to the collaborative mechanism among agents and the joint optimization of the value function. These allow for real-time identification of resource pressure and rapid, targeted adjustment.

Regarding performance variability, the latency variance for all tenants remains at a low level. The fluctuation range is under 1.2. This shows that the strategy not only ensures resource independence but also maintains consistency in service response. This is a key advantage of the multi-agent strategy over centralized scheduling models. It helps sustain stable system operation in complex environments.

In summary, the experimental results validate that the proposed strategy offers strong resource isolation and consistent service delivery in multi-tenant settings. The autonomy and collaboration of agents enable more effective resource governance in high-concurrency, multi-task cloud environments. This provides a practical solution for intelligent elastic management.

This paper also presents a robustness test of the multi-agent expansion strategy under burst traffic, and the experimental results are shown in Figure 3.

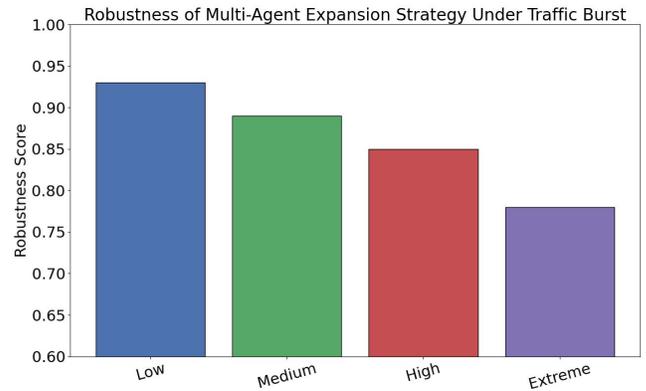

Figure 3. Robustness test of the multi-agent expansion strategy under burst traffic

Figure 3 illustrates that the multi-agent scaling strategy maintains high robustness under low to moderate burst traffic (scores of 0.93 and 0.89), while performance moderately declines under extreme conditions (down to 0.78) due to localized overloads and synchronization delays. Despite this, it consistently outperforms centralized scheduling, demonstrating strong adaptability and resilience in dynamic cloud environments.

V. CONCLUSION

This paper addresses the problem of elastic scaling in cloud resource management by proposing an intelligent scaling strategy based on a multi-agent system. The strategy builds a distributed agent cluster to achieve autonomous resource sensing, local decision-making, and coordinated scaling. It effectively overcomes the limitations of traditional methods, such as slow response, resource waste, and scheduling bottlenecks in complex and dynamic environments. Experimental results demonstrate that the proposed method achieves high resource utilization, strong service quality assurance, and system robustness under typical cloud scenarios, including multi-tenant environments and burst traffic disturbances. These results verify the practical feasibility and advantages of the method in elastic management tasks.

The core innovation of this study lies in the introduction of agent collaboration mechanisms and a state prediction module. These enhancements allow the system to learn scaling strategies dynamically from historical data and to anticipate resource demand trends. As a result, the system gains improved foresight and stability. The distributed nature of the multi-agent architecture significantly enhances scalability under large-scale deployments. This approach is well-suited for modern cloud platforms where high concurrency, heterogeneous tasks, and multi-tenant coexistence are common. The method also provides a novel path for the intelligent upgrade of cloud resource scheduling systems, offering theoretical and practical support for overcoming the limitations of centralized strategies. From an application perspective, the proposed multi-agent scaling mechanism can be widely applied to automated operations systems, intelligent edge computing platforms, and service-oriented microservice architectures. Its characteristics of high autonomy, adaptability, and scalability significantly improve dynamic resource scheduling, service stability, and operational cost control across different computing platforms. In addition, the method demonstrates strong capabilities in ensuring service quality and resource isolation. These features support the development of more intelligent, stable, and efficient cloud platforms, meeting the resource management demands of the industrial internet, smart cities, and large-scale online service systems.

## VI. FUTURE WORK

Future research may explore the integration of game-theoretic modeling among agents, adaptive communication mechanisms, and multi-objective scheduling optimization. These directions aim to enhance the system's adaptive collaboration in competitive resource environments. The findings may also be extended to cross-cloud or multi-edge cooperative scenarios, contributing to broader distributed computing architectures. By incorporating federated learning, security-enhanced strategies, or large language model-based decision modules, the intelligent level and deployment flexibility of multi-agent systems may be further advanced. This would promote cloud resource management technologies toward a higher stage of intelligent autonomy.